\documentclass[twocolumn,aps,prl,superscriptaddress,nofootinbib,showpacs,floatfix]{revtex4}

\usepackage{multirow}
\usepackage{epsfig}
\usepackage{amsmath}

\usepackage[normalem]{ulem}  
\usepackage[dvips]{color} 

\renewcommand\sout{\bgroup \color{red} \ULdepth=-.5ex \ULset}

\newcommand{\qbar}{\bar{q}}
\newcommand{\sbar}{\bar{s}}
\newcommand{\cbar}{\bar{c}}

\newcommand{\Ex}[2]{\ifmmode{#1\times10^{#2}}\else{$#1\times10^{#2}$}\fi}

\newcommand{\Psfig}[2]{\includegraphics[width=#1]{#2}}

\begin{document}
\title{Multi-quark hadrons from Heavy Ion Collisions}

\author{Sungtae Cho}\affiliation{Institute of Physics and Applied Physics, Yonsei
University, Seoul 120-749, Korea}
\author{Takenori Furumoto}\affiliation{Yukawa Institute for Theoretical Physics, Kyoto University, Kyoto 606-8502, Japan}\affiliation{RIKEN Nishina Center, Hirosawa 2-1, Wako, Saitama 351-0198, Japan}
\author{Tetsuo Hyodo}\affiliation{Department of Physics, Tokyo Institute of Technology, Meguro 152-8551, Japan}
\author{Daisuke Jido}\affiliation{Yukawa Institute for Theoretical Physics, Kyoto University, Kyoto 606-8502, Japan}
\author{Che Ming Ko}\affiliation{Cyclotron Institute and  Department of Physics and Astronomy, Texas A\&M
University, College Station, Texas 77843, U.S.A.}
\author{Su Houng~Lee}\affiliation{Institute of Physics and Applied Physics, Yonsei
University, Seoul 120-749, Korea}\affiliation{Yukawa Institute for Theoretical Physics, Kyoto University, Kyoto 606-8502, Japan}
\author{Marina Nielsen}\affiliation{Instituto de F\'{\i}sica, Universidade de S\~{a}o Paulo,
C.P. 66318, 05389-970 S\~{a}o Paulo, SP, Brazil}
\author{Akira Ohnishi}\affiliation{Yukawa Institute for Theoretical Physics, Kyoto University, Kyoto 606-8502, Japan}
\author{Takayasu Sekihara}\affiliation{Yukawa Institute for Theoretical Physics, Kyoto University, Kyoto 606-8502, Japan}\affiliation{Department of Physics, Graduate School of Science, Kyoto University, Kyoto 606-8502, Japan}
\author{Shigehiro Yasui}\affiliation{Institute of Particle and Nuclear Studies, High Energy Accelerator Research Organization (KEK), 1-1, Oho, Ibaraki 305-0801, Japan}
\author{Koichi Yazaki}\affiliation{Yukawa Institute for Theoretical Physics, Kyoto University, Kyoto 606-8502, Japan}\affiliation{RIKEN Nishina Center, Hirosawa 2-1, Wako, Saitama 351-0198, Japan}

\collaboration{ExHIC Collaboration}\noaffiliation

\date{\today}
\begin{abstract}
Identifying hadronic molecular states and/or hadrons with multi-quark components either with or without exotic quantum numbers is a long standing challenge in hadronic physics. We suggest that studying the production of these hadrons in relativistic heavy ion collisions offer a promising resolution to this problem as yields of exotic hadrons are expected to be strongly affected by their structures. Using the coalescence model for hadron production, we find that compared to the case of a non-exotic hadron with normal quark numbers, the yield of an exotic hadron is typically an order of magnitude smaller when it is a compact multi-quark state and a factor of two or more larger when it is a loosely bound hadronic molecule. We further find that due to the appreciable numbers of charm and bottom quarks produced in heavy ion collisions at RHIC and even larger numbers expected at LHC, some of the newly  proposed heavy exotic states could be produced and realistically measured in these experiments.
\end{abstract}


\maketitle

Finding hadrons with configurations other than the usual $q\bar{q}$ configuration for a meson and $qqq$ for a baryon is a long standing challenge in hadronic physics. In 1970's, the tetraquark picture~\cite{Jaffe76-1}  was suggested as an attempt to understand the inverted mass spectrum of the scalar nonet. At the same time, the exotic H dibaryon~\cite{Jaffe76} was proposed on the basis of the color-spin interaction. While  results from the long search for the H dibaryon in various experiments turned out to be negative, we are witnessing a renewed interest in this subject as the properties of several newly observed heavy states, including $D_{sJ}(2317)$~\cite{Aubert:2003fg} and $X(3872)$~\cite{Choi:2003ue}, cannot be properly explained within the simple quark model.

An important aspect in understanding a multi-quark hadron involves the discrimination between a compact multi-quark configuration and a loosely bound molecular configuration with or without exotic quantum numbers.  While the wave function of a loosely bound molecular configuration is dominantly composed of a bound state of well separated hadrons, the main Fock component of a compact multi-quark configuration typically has the size of a hadron, with little if any separable color singlet components. For a crypto-exotic state, one further has to distinguish it from a normal quark configuration.  For example, $f_0(980)$ and $a_0(980)$ could be either normal quark-antiquark states~\cite{RRY84}, compact tetra-quark states~\cite{Jaffe76-1} or weakly bound $K \bar{K}$ molecules~\cite{Weinstein82}.

Previously, discriminating between different configurations for a hadron relied on information about the detailed properties of the hadron and its decay or reaction rate~\cite{Close92}.  Moreover, searches for exotic hadrons have usually been pursued in reactions between elementary particles.  In this letter, we will show that measurements from  heavy ion collisions at ultrarelativistic energies can provide new insights into the problem and give answers to some of the fundamental questions raised above~\cite{Chen:2003tn,Chen07,Lee07}. In particular,  we focus on the yields of multi-quark hadrons in heavy ion collisions. To carry out this task, we first use the statistical model~\cite{Andronic:2005yp}, which assumes that the produced matter in relativistic heavy ion collisions is in thermodynamical equilibrium and is known to describe the relative yields of normal hadrons very well, to normalize the expected yields. We then use the coalescence model~\cite{Sato:1981ez}, which is based on the sudden approximation by calculating the overlap of the density matrix of the constituents in an emission source with the Wigner function of the produced particle, to take into account the effects of the inner structure of hadrons, such as angular momentum~\cite{KanadaEn'yo:2006zk} and the multiplicity of quarks~\cite{Chen07}. The coalescence model has been extensively used to study both light nucleus production in nuclear reactions~\cite{Chen:2003qj} and hadron production from the quark-gluon plasma produced in relativistic heavy ion collisions~\cite{Greco:2003xt,Fries:2003vb,Hwa:2003bn,Chen:2006vc}. In particular, it has successfully explained the observed enhancement of baryon production in the intermediate transverse momentum region~\cite{Adcox02,Abelov07} and the quark number scaling of the elliptic flow of identified hadrons~\cite{Adler03,Sorensen04} as well as the yield of recently discovered antihypertritons in heavy ion collisions at RHIC~\cite{atriton10}.

In the statistical model, the number of produced hadrons of a given type $h$ is given by~\cite{Andronic:2005yp}
\begin{align}
\label{Eq:Stat} N_h^\mathrm{stat} = & V_H \frac{g_h}{2 \pi^2}
\int_0^\infty \frac{p^2 dp}{\gamma_h^{-1}e^{E_h/T_H} \pm 1}
\end{align}
with $g_h$ being the degeneracy of the hadron, and $V_H~(T_H)$ the volume (temperature) of the source when statistical production occurs. The fugacity is $\gamma_h= \gamma_c^{n_c+n_{\bar{c}}} e^{(\mu_B B + \mu_s S)/T_H}$, where $B$ and $S$ are the baryon and strangeness numbers of the hadron with corresponding chemical potentials $\mu_B$ and $\mu_S$,  and $n_c (n_{\bar{c}})$ the number of (anti-)charm quarks. For central Au+Au (Pb+Pb) collisions at $\sqrt{s_{NN}}=200$ GeV (5.5 TeV) at RHIC (LHC), values for these parameters have been determined in Refs.~\cite{Chen:2003tn,Zhang:2008zzc} for particles in one unit of central rapidity in an expanding fire-cylinder model:  $V_H=1908~(5152)$ fm$^3$, $T_H=175$ MeV, $\mu_s=10~(0)$ MeV, and $\mu_B=20~(0)$ MeV. We fix  $\gamma_c=6.40~(15.8)$ by requiring the expected total charm quark number $N_c=3~(20)$ extracted from initial hard scattering at RHIC (LHC) to be equal to the sum of the yields of  $D$, $D^*$, $D_s$ and $\Lambda_c$ estimated in the statistical model. We note that all statistically produced hadrons from this fire-cylinder are essentially in the central unit rapidity.

In the coalescence model, the number of hadrons of type $h$ produced from the coalescence of $n$ constituents, based on harmonic oscillator wave functions for the hadron internal structure, is given by
\begin{align}\label{nquark}
N_h^{\rm coal} \simeq& g_h\prod_{j=1}^n \frac{N_j}{g_j}
\prod_{i=1}^{n-1}
\frac{(4\pi\sigma_i^2)^{3/2}}{V(1+2\mu_iT\sigma_i^2)} \left[
\frac{4\mu_i T\sigma_i^2}{3(1+2\mu_{i}T\sigma_{i}^{2})}
\right]^{l_i},
\end{align}
if we use the non-relativistic approximation, neglect the transverse flow of produced matter, and consider only the central unit rapidity as in Refs.~\cite{Chen:2003tn,Chen07}. In Eq.(\ref{nquark}), $g_i$ is the degeneracy of the $i$th constituent and $N_i$ its number taken to be $N_u$=245 (662) and $N_s$=150 (405), and $V=$1000 (2700)fm$^3$ for  RHIC (LHC)~\cite{Chen:2003tn};  $l_i$ is $0$ (1) for a $s(p)$-wave constituent; and $\sigma_i=1/\sqrt{\mu_i \omega}$ with $\omega$ being the oscillator frequency and $\mu_i$ the reduced mass defined by $\mu_i^{-1}=m_{i+1}^{-1}+(\sum_{j=1}^i m_j)^{-1}$ with $m_{u,d}(m_s)$=300 (500)MeV. Eq.(\ref{nquark}) is applicable if the number of constituent particles $N_j$ is greater than one, which is the case even for charmed and bottom particles in central heavy ion collisions at RHIC and LHC. Furthermore, Eq.~(\ref{nquark}) shows that the addition of a $s$-wave or $p$-wave $u/d$-quark with $i=1$ leads to the coalescence factor of about 0.360 or 0.093, respectively. Therefore, hadrons with more constituents are generally suppressed, and the $p$-wave coalescence is hindered with respect to the $s$-wave coalescence~\cite{KanadaEn'yo:2006zk}.

\begin{table*}[htdp]
\caption{List of multi-quark states.  For hadron molecules, the oscillator frequency  $\omega_\mathrm{Mol.}$ is fixed using the binding energy ((B)) or the inter-hadron distance ((R)). The $\omega_\mathrm{Mol.}$
for last two states is taken from corresponding two-body system ((T)).}
\begin{tabular}{c|cccc|c|c|c|c|c}
\hline \hline
\parbox[c]{1.5cm}{Pariticle}
& \parbox[c]{1.0cm}{$m$ (MeV)} & \parbox[c]{0.3cm}{$g$} &
\parbox[c]{0.5cm}{$I$} & \parbox[c]{0.8cm}{$J\pi$} &
\parbox[c]{1.3cm}{$2q/3q/6q$}
& \parbox[c]{1.5cm}{~\\[-0.5ex]$4q/5q/8q$\\[0.5ex]}
& \parbox[c]{1.5cm}{Mol.}
& \parbox[c]{1.5cm}{~\\[-0.3ex]$\omega_\mathrm{Mol.}$ (MeV)\\[0.5ex]}
& \parbox[c]{1.5cm}{decay mode}
\\
\hline $f_0(980)$  & 980& 1&0 &$0+$    & $q\qbar~(L=1)$  & $q\qbar
s\sbar$
        & $\bar{K}K$        & 67.8(B)  & $\pi\pi$ (strong decay) \\
$a_0(980)$  & 980& 3&1  &$0+$   & $q\qbar~(L=1)$  & $q\qbar
s\sbar$
        & $\bar{K}K$        & 67.8(B)      & $\eta\pi$ (strong decay) \\
$D_s(2317)$ &2317& 1&0  &$0+$   & $c\sbar~(L=1)$  & $q\qbar
c\sbar$
        & $DK$          & 273(B)   & $D_{s}\pi$ (strong decay) \\
$X(3872)$   &3872& 3&0  &$1+$   & -     & $q\qbar c\cbar$
        & $\bar{D}\bar{D}^*$    & 3.6(B)  & $J/\psi\pi\pi$ (strong decay) \\

\hline $\Lambda(1405)$ &1405& 2&0  &$1/2-$ & $qqs~(L=1)$     &
$qqqs\qbar$   & $\bar{K}N$    & 20.5(R)$-$174(B)
    & $\pi\Sigma$ (strong decay)\\
$\bar{K}KN$ &1920& 4&1/2&$1/2+$ & $ - $     & $qqqs\sbar~(L=1)$ &
$\bar{K}KN$   & 42(R)
    & $K\pi\Sigma$, $\pi \eta N$ (strong decay)\\
$\bar{D}N$  &2790& 2&0  &$1/2-$ & -     & $qqqq\cbar$   &
$\bar{D}N$    & 6.48(R)
    & $K^{+}\pi^{-}\pi^{-}+p$ \\
\hline $\bar{K}NN$ &2352& 2&1/2&$0-$   & $qqqqqs~(L=1)$  &
$qqqqqq\,s\qbar$& $\bar{K}NN$ & 20.5(T)-174(T)
    & $\Lambda N$ (strong decay) \\
$\bar{D}NN$ &3734& 2&1/2&$0-$   & -     & $qqqqqq\,q\cbar$  &
$\bar{D}NN$   & 6.48(T)
    & $K^{+}\pi^{-}+d$, $K^{+}\pi^{-}\pi^{-}+p+p$ \\
\hline \hline
\end{tabular}
\label{summary}
\end{table*}

In applying the coalescence model to multi-quark hadron production, we fix the oscillator frequencies by requiring the coalescence model to reproduce the reference \textit{normal} hadron yields in the statistical model. This leads to $\omega=550$ MeV for hadrons composed of light quarks. For hadrons composed of light and strange(charm) quarks, we fix  $\omega_s$ ($\omega_c$) to reproduce the yields of $\Lambda(1115)~(\Lambda_c(2286))$ in the statistical model.  For the $\Lambda_c(2286)$ yield, we include the feed-down contribution according to $N_{\Lambda_c(2286)}^\mathrm{stat,total}
=  N_{\Lambda_c(2286)}^\mathrm{stat}
     +N_{\Sigma_c(2455)}^\mathrm{stat}
    +N_{\Sigma_c(2520)}^\mathrm{stat}
    +0.67\times N_{\Lambda_c(2625)}^\mathrm{stat}$.
Fitting this yield to that calculated in the coalescence model, we obtain $\omega_c=385$ MeV for $m_{c}=1500$ MeV. Similarly, we get $\omega_s=519$ MeV from the $\Lambda(1115)$ yield after including the feed-down from the octet and decuplet states.

The yields for weakly bound hadronic molecules are estimated using the coalescence of hadrons at the kinetic freezeout point ($T_{F}=125$ MeV, $V_F=11322~(30569)$ fm$^3$ for RHIC (LHC)).  If the radius for hadronic molecules is known, the oscillator frequency $\omega$ can be fixed by $\omega= 3/(2\mu_1 \langle{r^2}\rangle)$ for the 2-body $s$-wave state.  If only the binding energy is given, we use the relation $\mathrm{B.E.} \simeq\hbar^2/(2\mu_1 a_0^2)$ and $\langle{r^2}\rangle \simeq a_0^2/2$, with $a_0$ being the $s-$wave scattering length, between the binding energy and the rms radius to obtain  $\omega=6 \times \mathrm{B.E.}$. For example, for $f_0(980)$,  $\omega_{f_0(980)}=67.8$ MeV using ${\rm B.E.}_{f_0(980)} = M_{K^\pm}+M_{K^0,\bar{K}^0}-M_{f_0(980)}=11.3$ MeV.  Table \ref{summary} summarizes the parameters and possible decay modes for a selection of multi-quark candidates as well as proposed states $\bar{K}KN$\cite{Jido-Enyo}, $\bar{K}NN$\cite{Akaishi}, $\bar{D}N$, and $\bar{D}NN$~\cite{Yasui:2009bz}.

\begin{table*}[htdp]
\caption{Yields in one unit of central rapidity
with oscillator frequencies $\omega=550$ MeV, $\omega_s=519$ MeV, and $\omega_c=385$ MeV.}
\label{results}
\begin{center}
\begingroup
\renewcommand{\arraystretch}{1.2}
\begin{tabular}{c|c|c|c|c|c|c|c|c}
\hline \hline & \multicolumn{4}{|c}{RHIC}
& \multicolumn{4}{|c}{LHC}\\
\cline{2-9} & 2q/3q/6q & 4q/5q/8q & Mol. & Stat. & 2q/3q/6q &
4q/5q/8q & Mol. & Stat.
\\
\hline
$f_0(980)$      &3.8, 0.73($s\sbar$)          &0.10           &13
&5.6
                &10, 2.0 ($s\sbar$)             &0.28           &36             &15             \\
$a_0(980)$      &11             &0.31           &40 &17
                &31             &0.83           &\Ex{1.1}{2}    &46             \\
$D_{s}(2317)$   &\Ex{1.3}{-2}   &\Ex{2.1}{-3} &\Ex{1.6}{-2}
&\Ex{5.6}{-2}
                &\Ex{8.7}{-2}   &\Ex{1.4}{-2}   &0.10           &0.35           \\
$X(3872)$       &---            &\Ex{4.0}{-5}   &\Ex{7.8}{-4}
&\Ex{2.9}{-4}
                &---            &\Ex{6.6}{-4}   &\Ex{1.3}{-2}   &\Ex{4.7}{-3}   \\
\hline \hline $\Lambda(1405)$ &0.81           &0.11 &1.8$-$8.3
&1.7
                &2.2            &0.29           &4.7$-$21       &4.2           \\
$\bar{K}KN$     &---             &0.019          &1.7 &0.28
                &---            &\Ex{5.2}{-2}   &4.2            &0.67           \\
$\bar{D}N$      &---            &\Ex{2.9}{-3}   &\Ex{4.6}{-2}
&\Ex{1.0}{-2}
                &---            &\Ex{2.0}{-2}   &0.28           &\Ex{6.1}{-2}   \\
\hline \hline $\bar{K}NN$     &\Ex{5.0}{-3}   &\Ex{5.1}{-4}
&0.011$-$0.24 &\Ex{1.6}{-2}
                &\Ex{1.3}{-2}   &\Ex{1.4}{-3}   &$0.026-0.54$   &\Ex{3.7}{-2}   \\
$\bar{D}NN$     &---            &\Ex{2.9}{-5}   &\Ex{1.8}{-3}
&\Ex{7.9}{-5}
                &---            &\Ex{2.0}{-4}   &\Ex{9.8}{-3}   &\Ex{4.2}{-4}   \\
\hline \hline
\end{tabular}
\endgroup
\end{center}
\end{table*}

The yields of states listed in Table~\ref{summary} are summarized in Table~\ref{results}. For example, possible configurations of the $f_0(980)$ could be an $s \bar{s}$ or a $u \bar{u}$ and $d \bar{d}$ state in addition to crypto-exotic  configurations discussed before.   For most of the states considered here, the coalescence yield from the compact multi-quark state is an order of magnitude smaller than that from the usual quark configuration as the coalescence of additional quarks are suppressed. Also, for the same hadronic state, the coalescence yield from the molecular configuration is similar to or larger than that from the statistical model prediction. The similarity in the yields from the statistical model and the coalescence model prediction for a  molecular configuration, despite the difference in the production temperatures $T_C$ and $T_F$, can be attributed to the larger size of the molecular configuration forming at lower temperature but at a larger volume; hence the ratio of volumes $\sigma_i^3/V$ is similar. The predicted appreciable yields of hadronic molecules in relativistic heavy ion collisions are in sharp contrast to those in high energy pp collisions, where molecular configurations with small binding energy are hard to produce, particularly at high $p_T$~\cite{Bignamini:2009sk}.  Our results do not change much if different forms of hadron wave functions are used, and the temperature, chemical potentials are varied in a reasonable range. Moreover, the correlated uncertainties in the number of charm quark and its fugacity largely cancel out in the studied ratios.

Our results also indicate that the yields of many multi-quark hadrons are large enough to be measurable in experiments. In particular, the heavy exotic hadrons containing charm or strange quarks can be produced at RHIC with appreciable abundance and even more so at LHC.  Moreover, since the newly proposed states with charm quark are below the strong decay threshold, the background of their weak hadronic decays could be substantially reduced through vertex reconstruction. Since the expected number of $D^0$ observed through the vertex detector is of order $10^5$ per month at LHC, even the ${\bar D}NN$ states is definitely measurable. Therefore, relativistic heavy ion collisions provide a good opportunity to search for multi-quark hadrons, and it may very well lead to the first observation of new multi-quark hadrons.

\begin{figure}[h]
\Psfig{7.5cm}{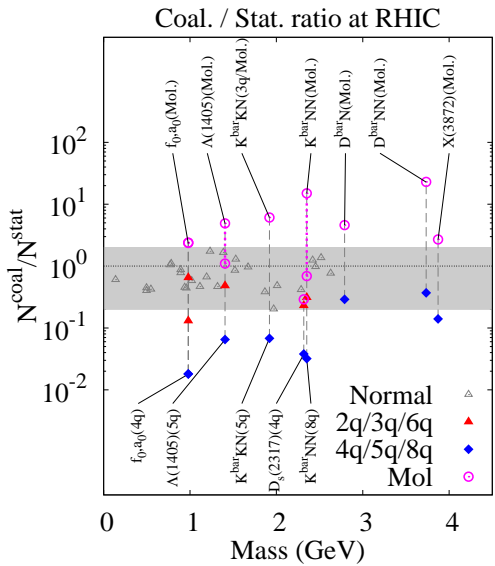}
\caption{(Color online) Ratio of hadron yields at RHIC in the coalescence model to those in the statistical model.}
\label{Fig:Mass}
\end{figure}

In Fig.~\ref{Fig:Mass}, we show the ratio $R_h$ of the yields at RHIC calculated in the coalescence model $N_h^\mathrm{coal}$ to those of the statistical model $N_h^\mathrm{stat}$ for the hadrons given in Table I. The grey zone within the range of $0.2 < R_h < 2$ denotes the range of the ratios for normal hadrons with $2q$ and $3q$ which are denoted by triangles inside the grey band.  The ratios for the crypto-exotic hadrons with usual $2q/3q$ configurations also fall inside the grey band.  The circles indicate the ratios obtained by assuming hadronic molecular configurations and are found to lie mostly above the normal band ($R_h > 2$).  Moreover, we find that
loosely bound extended molecules with larger size would be formed more abundantly. One typical example is $\Lambda(1405)$. Using the previous relation between the binding energy and the oscillator frequency $\omega$, we find a small size for $\Lambda(1405)$ ($\omega=174$ MeV) and a ratio $R_h=1.1$. A coupled channel analysis~\cite{Jido:2003cb,Hyodo:2007jq,Sekihara:2008qk} gives, however, a larger $\langle r^2 \rangle$, leading thus to a larger $R_h=4.9$.  The patterns shown in Fig.~\ref{Fig:Mass} also holds for LHC as the freezeout conditions are similar to those at RHIC.

As shown by  diamonds in Fig.~\ref{Fig:Mass}, the ratio $R_h$ is below the normal band ($R_h < 0.2$) when a hadron has a compact multi-quark configuration. In particular, for light quark configurations, these ratios are order of magnitude smaller than those of  normal hadrons or molecular configurations.  This is consistent with the naive expectation that the probability to combine $n$-quarks into a compact region is suppressed as $n$ increases. The tetraquark states of $f_0(980)$ and $a_0(980)$ are typical examples.  This suppression also applies to $5q$ states in multi-quark hadrons ($\Lambda(1405)$ and $\bar{K}KN)$ and the $8q$ state in $\bar{K}NN$. On the other hand, the yield of hadrons at higher transverse momenta is expected to be enhanced if they have multi-quark configurations since quark coalescence enhances the baryon/meson ratio at intermediate transverse momenta~\cite{Greco:2003xt,Fries:2003vb,Hwa:2003bn} as observed in experiments~\cite{Adcox02,Abelov07}.

We conclude from the above discussions that the yield of a hadron in relativistic heavy ion collisions reflects its structure and thus can be used as a new method to discriminate the different pictures for the structures of multi-quark hadrons. As a specific example, we consider $f_0(980)$. So far STAR has a preliminary measurement of $f_0(980)/\pi$ and $\rho^0/\pi$ from which we find $f_0(980)/\rho^0 \sim0.2$~\cite{STAR_f0}. Using the statistical model prediction for the yield of $\rho^0=42 $ leads to $f_0(980) \sim 8$.  Comparing this number to the numbers predicted for $f_0(980)$ in Table II, we find the data consistent with the $K \bar{K}$ picture.  Therefore, despite the quoted experimental error of around 50\%, the STAR data can be taken as an evidence that the $f_0(980)$ has a substantial $K\bar{K}$ component, and a pure tetraquark configuration can be ruled out for its structure.  Such conclusion could not be reached from analyzing the data for $f_0(980)\to2\gamma$~\cite{Close92,Pennington:2008xd}. Because of the large error bars in the STAR data, further experimental effort is highly desirable for putting an end to this controversial issue.  Similarly, efforts to measure the yields of other hadrons and newly proposed exotic states listed in Table I will provide new insights to a long standing challenge in hadronic physics.

\textit{Acknowledgements}
This work was supported in part by
the Yukawa International Program for Quark-Hadron Sciences at YITP, Kyoto University, the Korean BK21 Program and KRF-2006-C00011, KAKENHI (Nos. 21840026, 22105507 and 22-3389), the Grant-in-Aid for Scientific Research (Nos. 21105006 and 22105514), the global COE programs from MEXT, the U.S. National Science Foundation under Grant No. PHY-0758115, the Welch Foundation under Grant No. A-1358, and the CNPq and FAPESP. We thank the useful discussions with other participants during the YIPQS International Workshop on ``Exotics from Heavy Ion Collisions" when this work was started.

\end{document}